\documentclass[aps,pra,11pt,onecolumn,showpacs,superscriptaddress,floatfix,footinbib,subfigure]{revtex4}
\usepackage{amsmath}
\usepackage{latexsym}
\usepackage{amssymb}
\usepackage{amsopn}

\usepackage{graphicx,epstopdf,subfigure,color,hyperref}
\definecolor{Red}{rgb}{1,0,0}

\def\bra#1{\mathinner{\langle{#1}|}}
\def\ket#1{\mathinner{|{#1}\rangle}}
\def\braket#1{\mathinner{\langle{#1}\rangle}}

{\catcode`\|=\active
  \gdef\Braket#1{\begingroup
\mathcode`\|32768\let|\BraVert\left<{#1}\right>\endgroup}}
\def\BraVert{\egroup\,\mid\,\bgroup}

\def\braket#1#2{\mathinner{\langle{#1}|{#2}\rangle}}
\def\es{{\mathcal{S}}}
\def\md{{\mathcal{M}}}
\def\ea{{\mathcal{A}}}
\def\sm{{\es,\md}}

\def\sn{ \sigma_{ \hat{n} } }

\def\kb#1{\ket{#1}\bra{#1}}

\begin{document}
\title{Experimental realization of post-selected weak measurements on an NMR quantum processor}
\author{Dawei Lu}
\affiliation {Institute for Quantum Computing and Department of Physics \& Astronomy, University of Waterloo, Waterloo, Ontario N2L 3G1, Canada}
\author{Aharon Brodutch}
\email{aharon.brodutch@uwaterloo.ca}
\affiliation {Institute for Quantum Computing and Department of Physics \& Astronomy, University of Waterloo,  Waterloo, Ontario N2L 3G1, Canada}
\author{Jun Li}
\affiliation {Institute for Quantum Computing and Department of Physics \& Astronomy, University of Waterloo,  Waterloo, Ontario N2L 3G1, Canada}
\affiliation{Department of Modern Physics, University of Science
and Technology of China, Hefei, Anhui 230026, China}
\author{Hang Li}
\affiliation {Institute for Quantum Computing and Department of Physics \& Astronomy, University of Waterloo,  Waterloo, Ontario N2L 3G1, Canada}
\affiliation{Department of Physics, Tsinghua University, Beijing 100084, China}
\author{Raymond Laflamme}
\email{laflamme@iqc.ca}
\affiliation {Institute for Quantum Computing and Department of Physics \& Astronomy, University of Waterloo, Waterloo, Ontario N2L 3G1, Canada}
\affiliation {Perimeter Institute for Theoretical Physics, Waterloo, Ontario N2L
2Y5, Canada}
\date{\today}

\begin{abstract}
The ability to post-select the outcomes of an experiment is a useful theoretical concept and experimental tool. In the context of   weak measurements, post-selection can lead to surprising results such as complex weak values outside the range of eigenvalues.  Usually post-selection is realized by a projective measurement, which is hard to  implement in ensemble systems such as NMR.   We demonstrate the first experiment of a weak measurement with  post-selection on an  NMR quantum information processor. Our setup is used for measuring  complex weak values and weak values outside the range of eigenvalues.  The scheme for overcoming the problem of post-selection in an ensemble quantum computer is general and can be applied to any circuit-based implementation. This experiment  paves the way for studying and exploiting post-selection and weak measurements  in systems where projective measurements are hard to realize experimentally.

\end{abstract}
\maketitle
\section{Introduction}
Many fundamental experiments in quantum mechanics can be written and fully understood in terms of a quantum circuit \cite{Deutsch1985}. The current state of the art, however,  provides a poor platform for implanting these circuits on what can be considered a universal quantum computer. The limited number of  independent degrees of freedom  that can be manipulated efficiently limit the possible experiments. Many popular experiments involving  a small number of qubits  include some type of  post-selection. By post-selecting,  experimenters  condition their statistics only on those experiments that (will) meet a certain criteria such as the result of a projective measurement performed at the end of the experiment. Such conditioning produces a number of surprising effects  like  strange weak values \cite{Aharonov1988} and nonlinear quantum gates \cite{Lloyd2011}. Although they are sometimes described via quantum circuits, experiments  involving post-selection are usually implemented in a dedicated setup  which is not intended to act as a universal quantum processor. For example in current optical implementations  \cite{Lundeen2006,Rozema2012} the addition of more gates would require extra hardware.

Nowadays, implementations of quantum computing tasks in NMR architectures provide a good test-bed for up to 12 qubits  \cite{Negrevergne2006}.  Universal quantum circuits can  be implemented  on  mid-scale (5-7 qubit) NMR quantum computers with the development of high-fidelity control \cite{NMRRsoc}. However, the difficulty in performing  projective measurements poses a  handicap with the implementation of  circuits involving  post-selection. Here we will show how to overcome this  difficulty theoretically and experimentally, in the setting of weak measurements.

Weak measurements provide an elegant way to learn something about a quantum system $\es$  in the interval between preparation and post-selection \cite{Aharonov1991}. When the interaction between $\es$ and a measuring device $\md$ is weak enough, the back-action is negligible \cite{Tollaksen2010}. Moreover the effective evolution of the measuring device during the measurement is proportional to a \emph{weak value}, a complex number which is a function of the pre-selection, post-selection and the desired observable. These weak values allow us to make statements that would otherwise be in the realm of \emph{counterfactuals} \cite{Aharonov2002,Molmer2001}. Weak values have been interpreted  as complex probabilities \cite{Hofmann2013} and/or  \emph{element of reality} \cite{Vaidman1996} and, although somewhat controversial \cite{Peres1989, Aharonov1989, Leggett1989}, they are  used to describe  a number of fundamental issues in quantum mechanics including: Non-locality in the two slit experiment \cite{Tollaksen2010}, the trajectories of photons  \cite{Kocsis2011}, the reality of the wave function\cite{Lundeen2011}, Hardy's paradox\cite{Aharonov2002,Molmer2001,Lundeen2009}, the three box paradox \cite{Aharonov1991,Vaidman1996,Resch2004},   measurement precision-disturbance relations  \cite{Rozema2012,Lund2010} and the Leggett-Garg inequality \cite{Goggin2011,Williams2008,Groen2013}.  Recently the amplification effect associated with large real and imaginary weak values was  used in   practical schemes for precision measurements.  While these schemes  cannot  overcome the limits imposed by quantum mechanics, they can be used to improve precision under  various types of operational imitations  such as \emph{technical noise} \cite{Jordan2013}.

Experimental realizations of weak measurements  have so far been limited to optics with only a few recent exceptions  \cite{Shomroni2013, Groen2013}. In all cases post-selection was done using projective measurements that physically select events with successful  post-selection.  Optical experiments (e.g ref. \cite{Ritchie1991}) exploit the fact that one degree of freedom can be used as the system while another (usually a continuous degree of freedom) can be used as the measuring device.  Post-selection is  achieved  by filtering out photons that fail post-selection and the readout is done at the end only on the surviving systems. This type of filtering is outside the scope of ensemble quantum computers where all operations apart from the final ensemble measurement are unitary. One can overcome the difficulty by either  including a physical filter or by finding some way for post-selection using unitary operations. The former poses a technical challenge  as well as a conceptual deviation from the circuit model. In what follows we show how to perform the latter by resetting the measuring device each time post-selection fails. Our theoretical proposal is compatible with a number of  current implementation of quantum processors  such as such as  liquid and solid state NMR\cite{Negrevergne2006,Baugh2005,spatial,Gershenfeld1997},  electron spin qubits \cite{Sato2009,Kane1998} and rare earth crystal implementations  \cite{Ohlsson200271}.

 We begin by describing the weak measurement process for qubits and the theoretical circuit  for weak measurements on an ensemble quantum computer (Fig. \ref{circuit}). Next we describe our experiment in detail (Fig. \ref{molecule}). We demonstrate two properties of weak values associated with post-selection: weak values outside the range of eigenvalues and imaginary weak values (Figs. \ref{gweak}, \ref{thetaweak}). We conclude with a discussion of future applications


\section{Weak measurements}
The weak measurement  procedure (for qubits)  \cite{Brun2008}  involves a system $\es$ qubit  initially prepared in the state $\ket{\psi}^\es$ and a measuring device qubit  initially in the state $\ket{\md_i}^\md$ such that $\bra{\md_i}\sigma_z\ket{\md_i}=0$. The system and measuring device are coupled for a very short time  via the interaction  Hamiltonian  $H_i=\delta(t-t_i)g\sn^\es\sigma_z^\md$ with the coupling constant  $g<<1$  and $\sn=\hat{n}\cdot\vec{\sigma}$ a Pauli observable in the direction $\hat{n}$.  This is followed by a projective (post-selection) measurement on $\es$ with outcome $\ket{\phi}^\es$. Up to normalization we have

\begin{eqnarray}
\ket{\psi,\md_i}^\sm&&\rightarrow e^{-ig\sn^\es\sigma_z^\md}\ket{\psi,\md_i}^\sm  \nonumber\\
&&\rightarrow \cos(g)\langle {\phi}|{\psi}\rangle \ket{\phi,\md_i}^\sm-i\sin(g)\bra{\phi}\sn\ket{\psi}\sigma_z^\md\ket{\phi,\md_i}^\sm
\end{eqnarray}

The unnormalized state of $\md$ at the end is
\begin{equation}
\ket{\md_f}=\langle {\phi}|{\psi}\rangle[ \cos(g)\openone^\md-i\sin(g)\{\sn\}_w\sigma_z^\md]\ket{\md_i}^\sm
\end{equation}

where
\begin{equation}
\label{weakvalue}
\{\sn\}_w=\frac{\bra{\phi}\sn^\es\ket{\psi}}{\langle {\phi}|{\psi}\rangle}
\end{equation}
is  the weak value of $\sn$. We can now take the weak measurement approximation $|g\{\sn\}_w|<<1$ so that up to first order in $g$ we have
\begin{equation}
\ket{\md_f}\approx e^{-ig\{\sn\}_w\sigma_z^\md}\ket{\md_i}
\end{equation}

The measurement device is rotated by $g\{\sigma_{\hat{n}}\}_w$ around the $z$-axis.  Note that this is the actual unitary evolution (at this approximation), not an average unitary.  If we set the measuring device to be in the initial state $\frac{1}{\sqrt{2}}[\ket{0}+\ket{1}]$,  the  expectation value for $\sigma_{\hat{m}}^\md=\hat{m}\cdot\vec{\sigma}^\md$ will be
\begin{align}\label{readout}
\bra{\md_f}\sigma_{\hat{m}}^\md\ket{\md_f}&\approx\bra{\md_i}(\openone+ig\{\sn\}_w^*\sigma_z^\md)\sigma_{\hat{m}}^\md(\openone-ig\{\sn\}_w\sigma_z^\md)\ket{\md_i}\\&\approx\bra{\md_i}\sigma_{\hat{m}}^\md\ket{\md_i}+ig\bra{\md_i}\{\sn\}_w^*\sigma_z^\md\sigma_{\hat{m}}^\md-\{\sn\}_w\sigma_{\hat{m}}^\md\sigma_z^\md\ket{\md_i}\\
\label{ReIm}&=\bra{\md_i}\sigma_{\hat{m}}^\md\ket{\md_i}+igRe(\{\sn\}_w)\bra{\md_i}\sigma_z^\md\sigma_{\hat{m}}^\md-\sigma_{\hat{m}}^\md\sigma_z^\md\ket{\md_i}\nonumber\\&\;+gIm(\{\sn\}_w)\bra{\md_i}\sigma_z^\md\sigma_{\hat{m}}^\md+\sigma_{\hat{m}}^\md\sigma_z^\md\ket{\md_i}
\end{align}
Choosing $\hat{m}$ appropriately we can get  the real or imaginary part of the weak value.

The scheme above is based on the fact that $\es$ was post-selected in the correct state $\ket{\phi}$. Since this cannot be guaranteed in an experiment, we need to somehow disregard those events where post-selection fails.  As noted previously, the standard method for implementing the post-selection is by filtering out the composite $\sm$ systems that fail post-selection. While this is relatively simple in some  implementations  it  is not a generic property of a quantum processor. In particular post-selection is not implicit in the circuit model. The method to overcome this problem  is to reset the measuring device whenever post-selection fails.

\begin{figure}[h] \centering
\includegraphics[width=\columnwidth]{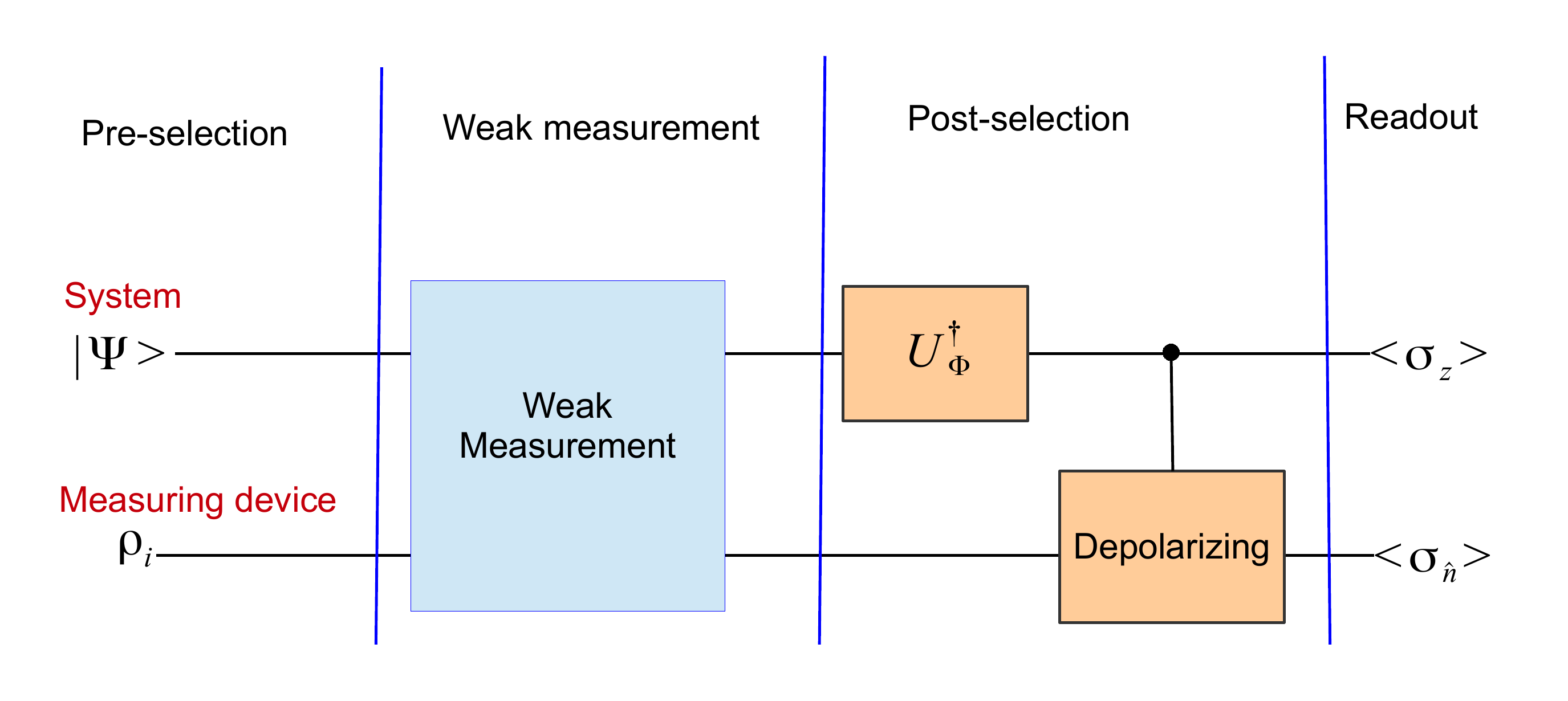}
\caption{(color online). Conceptual circuit for  post-selected weak measurements. The system $\es$  is prepared in the state $\ket{\psi}$, it then interacts weakly with the measurement device $\md$ (the weak measurement stage). Post-selection is performed by a controlled depolarizing channel in the desired basis, Eq \eqref{Cdepol}, decomposed into a rotation $U^\dagger_\phi$ followed by a contolled depolarizing in the computational basis. Finally at the readout stage we get the probability of post-selection $p_0$ from   $\langle\sigma_z^\es\rangle=2p_0-1$. The exception values $\langle\sn^\md\rangle$  are  shifted by a factor proportional to the weak value and decayed by a factor $p_0$. }\label{circuit}
\end{figure}

One way to implement the controlled reset  (see Fig. \ref{circuit})  is by using a controlled depolarizing gate with the following properties
\begin{align}\label{Cdepol}
\ket{\phi}\bra{\phi}^\es\otimes\rho^\md&\rightarrow \ket{\phi}\bra{\phi}^\es\otimes\rho^\md \nonumber\\
\kb{\phi^\perp}^\es\otimes\rho^\md &\rightarrow \kb{\phi^\perp}^\es\otimes\openone/2,
\end{align}
where  $\rho^\md$ is an arbitrary state and $\ket{\phi^\perp}^\es$ is the orthogonal state to the post-selection  $\braket{\phi^\perp}{\phi}=0$. This gate is not unitary and requires an ancillary system $\ea$. It can be  constructed  by setting $\rho^\ea=\openone/2$ and using a controlled-SWAP with the swap between $\ea$ and $\md$. To keep the control in the computational basis, we  preceded the controlled-SWAP gate by $U_\phi^{\es\dagger}$ where $U_\phi\ket{0}=\ket{\phi}$.  At the readout stage we  need to measure the observable $\kb{0}$ which will give the probability of post-selection $p_0$ as well as $\langle\sigma_{\hat{m}}^\md\rangle$. Finally we divide $\langle\sigma_{\hat{m}}^\md\rangle$  by $p_0$ to obtain the  weak value. If we are  only interested in reading out the real or imaginary part of the weak value we can further simplify this operation by a controlled dephasing in a complementary basis to $\sigma_{\hat{m}}^\md$. This allows us to replace the controlled-SWAP by a controlled-controlled-phase with $\md$ as the target and $\ea,\es$ as the control (see Fig. \ref{molecule}b). In our experimental setup  we  used the latter (simplified) approach.  A controlled-controlled-$\sigma_z$ was used in the post-selection for measuring real weak values and controlled-controlled-$\sigma_x$ was used for  complex weak values.

\section{Experimental Implementation}

\begin{figure}[t] \centering
\includegraphics[width=0.85\columnwidth]{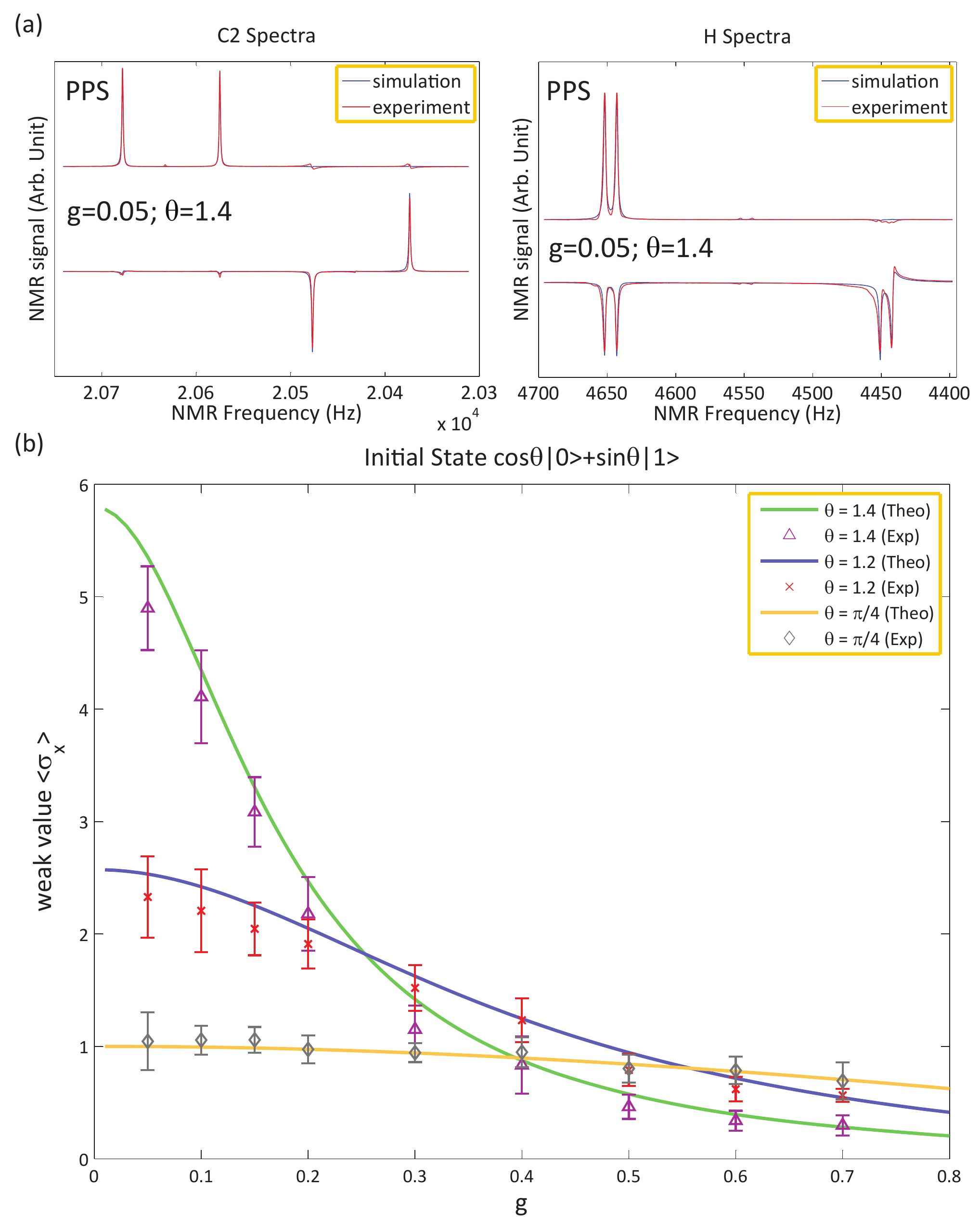}
\caption{(color online). (a) The spectra for each qubit (bottom spectrum) was compared with the reference PPS  (top spectrum) to obtain the expectation values $\langle\sigma_y^\md\rangle$ on C2 and $\langle\sigma_z^\es\rangle$ on H.   To observe the signal  on H, we  applied $\pi/2$ readout pulses after the sequence.  The spectra in the figure are for  $g=0.05$ and $\theta = 1.4$, the simulated spectra (blue) fit well with the experimental one (red).   (b) Weak values for various initial states, $\cos(\theta)\ket{0}^\es+\sin(\theta)\ket{1}^\es$,  were calculated using Eq \eqref{weakvalue}  and  compared the theoretical predictions  as a function of the measurement strength  from  $g=0.05$  to $g=0.7$.   The solid curves are theoretical predictions without the weak measurement approximation.  When the overlap between the pre and post-selection, $\cos(\theta)$  is large enough we  get very close to the asymptotic $g\rightarrow 0$  value at $g\ge0.05$.  However for $\theta=1.4$ the interaction was not weak enough at $g=0.05$.  The error bars are plotted by repeating each experiment four times. At low post-selection probabilities  we see observed values decrease  due to decoherence.}\label{gweak}
\end{figure}

\begin{figure}[h] \centering
\includegraphics[width=\columnwidth]{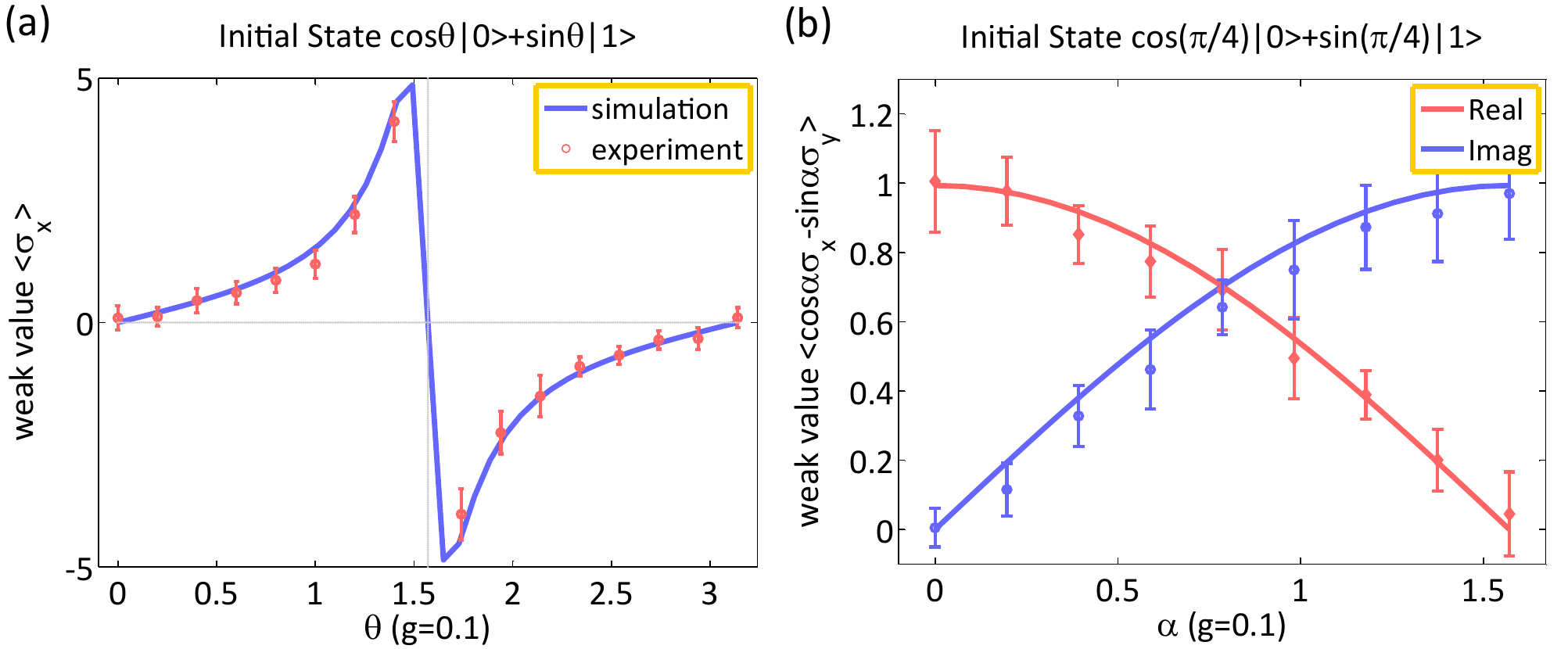}
\caption{(color online). (a) Experimental weak values $\{\sigma_x\}_w$ as a function of the initial state $\cos(\theta)\ket{0}+\sin(\theta)\ket{1}$  for a fixed post-selection $\ket{0}$  at $g=0.1$. The theoretical  (blue) curve is plotted for $g=0.1$  without the weak measurement approximation, and the experimental data is denoted by red circles. As expected at  $\pi/4<\theta<\pi-\pi/4$ we measure a relative shift outside the range of eigenvalues, $-1\le\langle\sigma_x\rangle\le1$. As $\theta$ approached $\pi/2$ the signal was too weak to measure due to the low probability of post-selection.  (b) Experimental result of measuring both the real and imaginary parts of the weak value $\{\cos(\alpha)\sigma_x-\sin(\alpha)\sigma_y\}_w$ at $g=0.1$ and $\theta = \pi/4$ as a function of $\alpha$.  At these values it was possible to get the full range of weak values with an absolute magnitude of 1.}\label{thetaweak}
\end{figure}

 We conducted three types of experiments.  In each  experiment we prepared $\es$   in  an initial state on the $x-z$ plane, $\cos(\theta)\ket{0}^\es+\sin(\theta)\ket{1}^\es$, the weak measurement observable was a Pauli operator  in the $x-y$ plane, $\cos(\alpha)\sigma_x^\es+\sin(\alpha)\sigma_y^\es$ and the post-selection was always the $\ket{0}^\es$ state. In the first experiment (Fig. \ref{gweak}b) we set $\alpha=0$ to make a measurement of $\sigma_x$ and varied the coupling strength from $g=0.05$  to $g=0.7$. We used three different initial states: $\theta=\pi/4$ - corresponding to a measurement where the weak value coincides with the result of a projective ($g=\pi/4$) measurement, $\theta=1.2$ -  where we were able to observe a large weak value,$\{\sigma_x\}_w=2.57$, at $g=0.05$ ,  and $\theta=1.4$ - where the weak value (first order) approximation is off by more than $10\%$ at $g=0.05$.  Next we kept the coupling constant at $g=0.1$ and varied over $\theta$  to observe real weak values both inside and outside the range of eigenvalues $[-1,1]$ (Fig. \ref{thetaweak}a). Finally we measured complex weak values with an absolute magnitude of 1 by keeping the initial state constant, $\theta=\pi/4$, and varying over the measurement direction $\alpha$ at $g=0.1$ (Fig. \ref{thetaweak}b).

\begin{figure}[h] \centering
\includegraphics[width=\columnwidth]{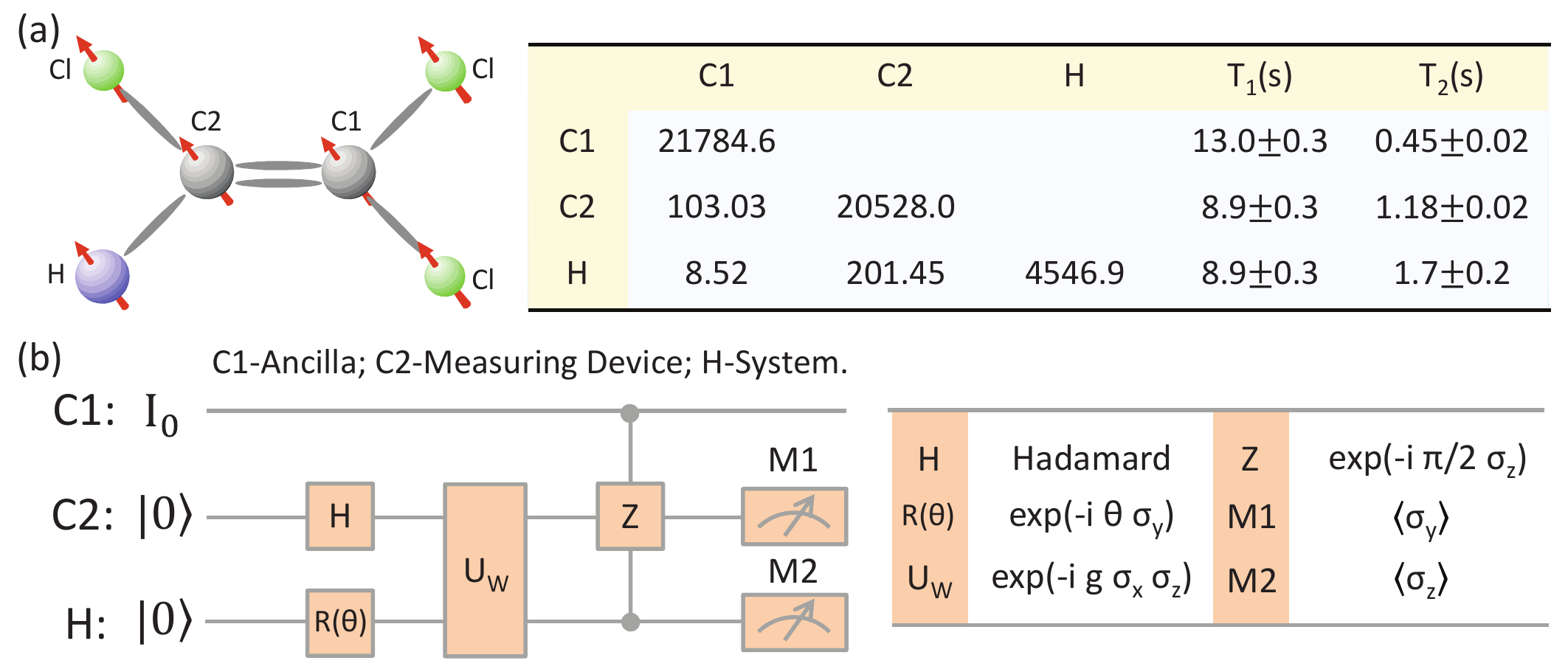}
\caption{(color online). (a) Experimental implementation of a weak measurement in NMR using trichloroethylene in which the two $^{13}$C and one $^{1}$H spins form a 3-qubit sample. In the parameter table  the diagonal elements are the chemical shifts (Hz), and the off-diagonal elements are scalar coupling strengths (Hz).  T$_1$ and T$_2$ are the relaxation and dephasing time scales.  (b) The quantum network used to realize the post-selected weak measurement in the experiment with  C1 as the ancilla $\ea$, C2 as the measuring device $\md$, and H as the system $\es$.  Post-selection of $\ket{0}^\es$ was  achieved using the controlled-controlled-Z gate so that $\md$ is dephased when post-selection fails.  The final measurements give the expectation values $\langle\sigma_z^\es\rangle$ and $\langle\sigma_y^\md\rangle$ which are used to calculate the weak value via  Eq \eqref{weakvalue}.  }\label{molecule}
\end{figure}

 All experiments were conducted on a Bruker DRX 700MHZ spectrometer at room temperature. Our 3-qubit sample  was  $^{13}$C labeled trichloroethylene (TCE) dissolved in d-chloroform. The structure of the molecule is shown in Fig. \ref{molecule}a, where we denote C1 as qubit 1, C2 as qubit 2, and H as qubit 3. The internal Hamiltonian of this system can be described as
\begin{eqnarray}\label{Hamiltonian}
\mathcal{H}=&&\sum\limits_{j=1}^3 {\pi \nu _j } \sigma _z^j  + \frac{\pi}{2}(J_{13}\sigma _z^1 \sigma _z^3+J_{23}\sigma _z^2 \sigma _z^3) \nonumber\\
&&+ \frac{\pi}{2}J_{12} (\sigma _x^1 \sigma _x^2+\sigma _y^1 \sigma _y^2+\sigma _z^1 \sigma _z^2),
\end{eqnarray}
where $\nu_j$ is the chemical shift of the $j$th spin and $J_{ij}$ is the scalar coupling strength between spins $i$ and $j$. As the difference in frequencies between C1 and C2 is not large enough to adopt the (NMR \footnote{We distinguish the term weak coupling as used in NMR from the term weak interaction used for the weak measurement}) weak J-coupling  approximation \cite{nmrreview}, these two carbon spins are treated in the strongly coupled regime. The parameters of the Hamiltonian are obtained by iteratively fitting the calculated and observed spectra through perturbation, and shown in the table of Fig. \ref{molecule}a.

We label C1 as the ancilla $\ea$, C2 as the measuring device $\md$, and H as the system $\es$ (Fig. \ref{molecule}b). Each experiment can be divided into four parts: (A) Pre-selection: Initializing the ancilla to the identity matrix $\openone/2$, the measurement device to $1/\sqrt{2}(|0\rangle+|1\rangle) $, and the system to $cos\theta|0\rangle+sin\theta|1\rangle$. (B) Weak measurement: Interaction between the measuring device and system, denoted by U$_w$ in the network. (C) Post-selection of the system in the state $|0\rangle$. (D) Measurement:  $\langle \sigma_y^\md \rangle$  on the measuring device ($\langle \sigma_z^\md \rangle$ for the imaginary part) and $\langle \sigma_z ^\es \rangle$ on the system. The experimental details of the four  parts above are as follows.

 (A) Pre-selection:  Starting from the thermal equilibrium state, first we excite the ancilla C1 to the transverse field by a $\pi/2$ rotation, followed with a gradient pulse to destroy  coherence leaving   C1 in the maximally mixed state $\openone/2$. Next we create the pseudopure state (PPS) of   C2 ($\md$) and  H ($\es$) with deviation $\openone \otimes \kb{00}$ using the spatial average technique \cite{spatial}.  For traceless observables and unital evolution the identity part of the PPS can be treated as noise on top of a pure state.  The spectra of the PPS followed by $\pi/2$ readout pulses are shown in the upper part of Fig. \ref{gweak}a. The readout pulses are applied on C2 (left figure) and H (right figure), respectively. Two peaks are generated in the PPS spectra because C1 is in the maximally mixed state $\openone/2$, and these two peaks are utilized as the benchmark for the following experiments. Finally  we apply one Hadamard gate on C2 and one $R_y(\theta) = e^{-i\theta\sigma_y}$ rotation on H. At the end of this procedure the state is
 \begin{eqnarray}\label{pps}
 \rho_{ini}= \openone\otimes \frac{1}{2} (|0\rangle + |1\rangle) (\langle 0 | + \langle 1 |) \otimes  (\mathrm{cos} \theta|0\rangle + \mathrm{sin}\theta|1\rangle)(\mathrm{cos}\theta\langle 0 | + \mathrm{sin}\theta\langle 1 |).
 \end{eqnarray}

 (B) Weak measurement:  The unitary operator to realize the weak measurement
 \begin{equation}\label{uw}
U_w=e^{-ig\sigma_{\hat{n}}^2 \sigma_z^3}.
 \end{equation}
 can be simulated by the interaction term $\sigma_z^2\sigma_z^3$ between C2 and H using the average Hamiltonian theory \cite{ernstbook}. However, since the internal Hamiltonian contains a strongly coupling term and the refocusing scheme requires the  WAHUHA-4 sequence \cite{wahuha} we  adopted the gradient ascent pulse engineering (GRAPE) technique \cite{grape1,grape2}  to improve the fidelity (see discussion below).

 (C) Post-selection: In order to mimic the post-selection of $|0\rangle$ on the system spin H in NMR, we introduce an ancilla qubit C1 in the maximally mixed state $\openone$. The controlled resetting noise operation is  a controlled-controlled-$\sigma_z$ gate
 \begin{equation}\label{postselection}
\openone\otimes \openone \otimes \openone - |1\rangle \langle 1 |  \otimes  |1\rangle \langle 1 |\otimes \openone + |1\rangle \langle 1 | \otimes |1\rangle \langle 1 | \otimes \sigma_z.
 \end{equation}
 If post-selection is successful, the measurement device will point to the weak value. Otherwise the measurement device will become dephased and the expectation value $\left<\sigma_y^2\right>$ will be reset to 0.  For the measurement of complex weak values we used a standard Toffoli and applied the same reasoning.

(D) Measurement: Finally we measure the expectation value $\langle \sigma_y^2 \rangle$ on C2 and $\langle \sigma_z^3 \rangle$ on H, to calculate the weak value by the expression
 \begin{equation}\label{weakvalue}
Re(\{ \sigma_x \}_w) \approx \frac{\langle \sigma_y^2 \rangle}{ g(\langle \sigma_z^3 \rangle+1)}.
 \end{equation}
For the imaginary part we similarly use $\left<\sigma_z^2\right>$.

Since the timescales for the experiment are much shorter than $T_1$ the evolution is very close to unital.

In the experiment, the $\pi/2$ rotation of H is realized by the hard pulse with a duration of 10 $\mu$s. All the other operations are implemented through GRAPE pulses to achieve high-fidelity control. These GRAPE pulses are designed to be robust to the inhomogeneity of the magnetic field, and the imprecisions of the parameters in the Hamiltonian. For the two selective $\pi/2$ excitations on C1 and C2, the GRAPE pulses are generated with the length 1.5 ms, segments 300 and fidelity over 99.95\%. These two pulses are only used for the preparation and observation of PPS. Besides these two GRAPE pulses, the PPS preparation involves another GRAPE pulse of 10 ms and 99.99\% fidelity for the creation of two-body coherence. The top spectra in Fig. \ref{gweak}a are  the  PPS both on C2 and H, which are used as the benchmark for the following experiments.

The main body of the network shown in Fig. \ref{molecule}b, including the pre-selection, weak measurement and post-selection, is calculated by a single GRAPE pulse. Since we have to alter the initial state by $\theta$ and interaction strength $g$ in the experiment, we have used different GRAPE pulses to implement the experiments with different group of parameters. All of these GRAPE pulses have the same length: 20 ms, and fidelity over 99.98\%.

There are two essential advantages in utilizing the GRAPE pulses: reducing the error accumulated by the long pulse sequence if we directly decompose the network, and reducing the error caused by  decoherence. Because the $U_w$ evolution is supposed to be very ``weak'', the intensity of the output signal is quite small. Moreover, the calculated weak values are more sensitive to the intensity of the signal, because the small $g$  in the denominator (Eq \ref{weakvalue}) will amplify the errors in the experiment. Therefore, we need to achieve as accurate coherent control as we can to obtain precise experimental results. The direct decomposition of the original network requires many single rotations, as well as a relatively long evolution time. For example, an efficient way to decompose the controlled-controlled-$\sigma_z$ gate is by six CNOT gates and several one qubit gates \cite{NielsenandChuang}, which requires more than 50 ms in our experimental condition. Therefore, using \emph{short} GRAPE pulses of \emph{high fidelity} we were able to decrease the potential errors  due to decoherence and imperfect implementation of the long pulse sequence.  Fig. \ref{gweak}a shows  the experimental spectra compared with the simulated one in the case $g=0.05$ and $\theta = 1.4$. The top plots are the reference (PPS) spectra while the bottom are the spectra measured at the end of the experiment. The predicted value of $\langle \sigma_y^2 \rangle$ in this case is only 1.67\%. However, since the experimental spectra are very close to the simulated one, we can obtain a good result after extracting the data. The effects of decoherence in the case of small post-selection probabilities values can be seen in Fig. \ref{gweak}b where the experimental values  are consistently lower than the theoretical predictions. Since decoherence reduces the expectation values the results of   Eq~\eqref{weakvalue} will be more sensitive to decoherence as $\langle\sigma_z^3\rangle\rightarrow - 1$.

In the first experiment we varied the interaction strength   $g=0.05$ to $g = 0.7$ for $\alpha=0$ and three different initial states $\theta = 1.4$, $\theta = 1.2$, and $\theta =\pi/4$. We obtained the weak value through a final measurement of  $\langle \sigma_y^2 \rangle$ on the measuring device C2 and $\langle \sigma_z^3 \rangle$ on the system H, respectively. The weak value $\{ \sigma_x \}_w$  was calculated through Eq \eqref{weakvalue}, with the result shown in Fig. \ref{gweak}b. The error bars were  obtained by repeating the experiment  four times. The large error bars in the weak values at small values of  $g$ and low post-selection probabilities  are a result of imperfect calibration of the low experimental signals obtained at that range.

In the second experiment  we studied  the behavior of measured weak values as a function of the initial state parameter $\theta$ at $g=0.1$ and $\alpha=0$. The theoretical weak value should be  $\tan(\theta)$ (at $g\rightarrow0$). We compared our results with a theoretical curve at $g=0.1$  (Fig. \ref{thetaweak}a). As expected this curve diverges from the weak value of $\tan(\theta)$ as the overlap between the pre and post-selection vanishes   and second order terms become more dominant. The experimental data matches with the smooth part of the curve. When $\theta$ is very close to  $\pi/2$, we were unable to measure the extremely low NMR signals  due to the signal-to-noise ratio(SNR) issues, moreover decoherence effects become more prominent at these values (see discussion above).

In the final experiment we  measured both the real part and imaginary parts of the weak value (Fig. \ref{thetaweak}b).  We set $\theta = \pi/4$ and $g=0.1$, and changed $\alpha$, the observable of the measuring device C2. The expected weak values have an absolute magnitude of 1 and the  overlap between the pre and post-selected state is $1/\sqrt{2}$.  To measure the imaginary part, we replaced the controlled-controlled-$\sigma_z$ gate in Fig. \ref{molecule}b with a controlled-controlled-$\sigma_x$ gate, and measured the expectation value $\langle\sigma_z^2\rangle$ on the measuring device C2. The other parts of the experiments remain the same. In  Fig. \ref{molecule}b we can see the real part and imaginary part of the weak value along with the parameter $\alpha$. The results match well with the theory.

\section{Conclusions and outlook}
Post-selection is a useful and interesting conceptual and practical tool. Its experimental implementations have so far  been limited to dedicated quantum devices. Here we showed that this paradigm can be realized in a more general setting of a quantum circuit with unitary gates. We  implemented a weak measurement where shifts corresponding to weak values outside the range of eigenvalues ($[-1,1]$  in the case of Pauli observables) are an artifact of non trivial post-selection.

Our experiment involved a simple 3-qubit system and we were able to demonstrate the measurement of large (Fig  \ref{gweak}b, \ref{thetaweak}a) and imaginary (Fig. \ref{thetaweak}b) weak values.  We observed  a shift of $2.33g$ with an accuracy of $\pm0.36g$  at $\theta=1.2$ for $g=0.05$    compared with the theoretical weak value of  $2.57g$ (at $g\rightarrow 0$). The largest relative shift observed was  $4.90g\pm  0.37g$ for  $\theta=1.4$  at   $g=0.05$ compared with a theoretical value of  $5.8g$. At this value the  weak measurement approximation would require the interaction to be an order of magnitude smaller.   For complex weak values we were able to demonstrate  the full spectrum of complex weak values with unit  absolute magnitude.

Our scheme has the advantage that it can be extended to systems with more qubits without significant changes.  Implementations on a four qubit system will allow the first fully quantum  implementation of the three box paradox  and further extensions will allow more intricate experiments such as the measurement of the wave-function and   measurement-disturbance relations which have so far been limited to optical implementations, often with classical light. The reasonably large number of qubits  that can be manipulated   in NMR systems will also allow more intricate experiments that  are not possible in optics.  It remains an open question whether our techniques can be used for precision measurements in the same way as they are used in optics.  Given that this is the first implementation of weak measurements in NMR there is still much to be explored.

\section*{}
We thank Marco Piani and Sadegh Raeisi for comments and discussion. This work was supported by  Industry Canada, NSERC  and CIFAR. J. L. acknowledges National Nature Science Foundation of China, the CAS, and the National Fundamental Research Program 2007CB925200.


\end{document}